\documentstyle[12pt]{article}

\documentstyle{titlepage}
\title{IS SUPERLUMINAL MOTION IN RELATIVISTIC BOHM'S THEORY OBSERVABLE?}
\author{\bf ALI\ SHOJAI$^*$\ \&\ MEHDI\ GOLSHANI$$\\
Department of Physics, Sharif University of Technology\\P.O.Box 11365-9161 Tehran, IRAN\\
and\\Institute for Studies in Theoretical Physics and Mathematics,\\P.O.Box 19395-5531, Tehran, IRAN\\
$^*$Email: SHOJAI@PHYSICS.IPM.AC.IR\\}
\date{}
\begin{document}
\begin{bf}
\maketitle
\vspace{1cm}
\begin{center}
{\Large IS SUPERLUMINAL MOTION IN RELATIVISTIC BOHM'S THEORY OBSERVABLE?}\\
{\bf A. Shojai \& M. Golshani}
\end{center}
\vspace{0.5cm}
\begin{center}
{\bf ABSTRACT}
\end{center}
{\it We show that the problem of superluminal motion in causal, particle interpretation
of bosonic fields is not observable at macroscopic distances.}
\vspace{1.5cm}
\section{INTRODUCTION}
\hspace{0.5cm}In extending Bohm's causal version of
quantum mechanics to the relativistic domain, one faces with a serious difficulty.
For bosonic wave functions, both space-like four-momenta as well as time-like ones
are possible. This shows that if one accepts the causal, particle interpretation
of bosonic wave functions, then, one must deal with particles moving sometimes 
faster than light. This objection has lead many authors to argue that a particle
interpretation of bosonic fields is not possible. They are fields rather than particles --
in contrast to the fermionic case in which a causal, particle interpretation 
is acheivable. 
\par
This does not seems to us a reasonable argument. First, because it can be shown
that even for bosonic fields it is possible to construct a time-like current 
four-vector.[1] Second, we show in
this letter that at macroscopic separations (i.e. separations larger than the
de-Broglie wavelength of the particle)
the velocity of the particle 
is less than that of the light -- at least under some conditions.
Bohm has already mentioned this posibility[2]:
\\
{\it .... it is necessary to assume a basically non-Lorentz invariant
theory for the individual particles. Nevertheless this theory becomes
Lorentz invariant where nonlocal connections can be neglected. From this
we conclude that the manifest level of ordinary large scale exprience
will be covariant in its behaviour. .... the statistical laws of the 
quantum theory are covariant. We emphasize again that there is Lorentz 
invariance in all of the domains of particle theory that have thus far been 
investigated experimentally, but these do not necessarily invalidate
our assumptions concerning the underlying level in which the order of
succession is unique.\/} 
\\
In this paper we shall demonstrate this fact for bosons.
\section{Superluminal motion in the bosonic fields}
\hspace{0.5cm}First, we briefly review 
what happens in making a causal, particle interpretation of the Klein-Gordon
equation. It can be easily shown that the Klein-Gordon equation
\begin{equation}
(\Box + m^2)\Phi=0
\end{equation}
is mathematically equivalent to the continuity equation
\begin{equation}
\partial _{\mu}(R^2 \partial ^{\mu}S)=0
\end{equation}
and the Hamilton-Jacobi equation
\begin{equation}
\partial _{\mu}S \partial ^{\mu}S=M^2
\end{equation}
where 
\begin{equation}
M^2=m^2+\frac{\Box R}{R}
\end{equation}
\begin{equation}
R=\sqrt{\Phi \Phi ^*}
\end{equation}
\begin{equation}
S=-\frac{i}{2}ln\left (\frac{\Phi}{\ \Phi ^*} \right )
\end{equation}
The four-momentum of the particle may be identified as $-\partial _{\mu}S$
and the mass of the prticle as $M$. Since $M^2$ is not positive definite, 
one may encounter with space-like four-momenta. To show such a possibility,
we consider the following solution of the $1+1$ dimensional Klein-Gordon
equation[3]
\begin{equation}
\Phi=N [exp(-imt)+Aexp(-i\omega t +ikx)]
\end{equation}
where 
\begin{equation}
\omega ^2 -k^2=m^2
\end{equation}
and $A$ is a real constant, and $N$ is the normalization factor. The energy 
and momentum of a particle guided by this wave function is
\begin{equation}
E=-\frac{\partial S}{\partial t}=\frac{1}{R^2}\left \{ m+A^2\omega +A(m+\omega )cos[(m-\omega )t-kx] \right \}
\end{equation}
\begin{equation}
P=\frac{\partial S}{\partial x}=\frac{1}{R^2}Ak \left \{ A+cos[(m-\omega )t-kx] \right \}
\end{equation}
The four-momentum is in many cases space-like, e.g. for $A=2/ \omega$, $\omega >2$ and  
$cos[(m-\omega )t-kx]=-1$. The velocity of the particle is
\begin{equation}
\frac{dx}{dt}=\frac{E}{P}=Ak\frac{A+cos[(m-\omega )t-kx]}{m+A^2\omega +A(m+\omega )cos[(m-\omega )t-kx]}
\end{equation}
Defining
\begin{equation}
\eta =(m-\omega )t-kx
\end{equation}
one has
\begin{equation}
\frac{d\eta}{dt}=\frac{m^2-A^2-(2\omega ^2-m^2-m \omega )-m\omega}{m+A^2\omega +A(m+\omega )cos\eta}
\end{equation}
equation (11) clearly shows that the velocity may occasionally be larger than unity.
This fact, is shown in figure for a typical trajectory dictated by (11).
But we claim that on the average the velocity is less than one as can be seen 
in the figure and also by averaging the relation (12)
\begin{equation}
<\frac{d\eta}{dt}>=0
\end{equation}
So
\begin{equation}
<\frac{dx}{dt}>=|\frac{m-\omega}{k}|=\sqrt{\frac{\omega -m}{\omega +m}}<1
\end{equation}
Therefore on the average we can not see any faster than light motion
for this particular solution. In the next section we shall prove this fact under some
general conditions.
\section{The average  velocity of a particle 
         in  the causal, particle interpretation 
         of  the Klein-Gordon equation}
\hspace{0.5cm}Now, we consider the general solution of the Klein-Gordon
equation 
\begin{equation}
\Phi=\int d^3k f(\vec{k})exp(-i\omega t+i\vec{k}\cdot \vec{x})
\end{equation}
where
\begin{equation}
\omega ^2 -|\vec{k}|^2 =m^2
\end{equation}
We show, by two methods, that the observed velocity is less 
than unity under suitable conditions.
\par
(a)--From the relation (5) and (6) we have for the solution (15)
\begin{equation}
\partial _{\mu}S=\frac{1}{2R^2}\int d^3k d^3k' f(\vec{k}) f^*(\vec{k'}) exp[-i(\omega -\omega ')t+i(\vec{k}-\vec{k'})\cdot \vec{x}]\{k_{\mu}+k'_{\mu}\}
\end{equation}
\begin{equation}
R^2=\int d^3k d^3k' f(\vec{k}) f^*(\vec{k'}) exp[-i(\omega -\omega ')t+i(\vec{k}-\vec{k'})\cdot \vec{x}]
\end{equation}
Now suppose that the k-space wave function $f(\vec{k})$ is localized in the region $|\vec{k}|\le \kappa$,
and assume that we are at a distance $x$, larger than $1/\kappa$. The integrals in
(17) and (18) then exist only when $\vec{k}\sim \vec{k'}$. Otherwise, the integrand 
oscilates rapidly and on the average leads to zero.
Thus we have
\begin{equation}
\partial _{\mu}S\rightarrow \frac{1}{R^2}\int d^3k  |f(\vec{k})|^2 k_{\mu}
\end{equation}
\begin{equation}
R^2 \rightarrow \int d^3k  |f(\vec{k})|^2
\end{equation}
The velocity is then 
\begin{equation}
\frac{dx_i}{dt}=-\frac{\ \partial S/\partial x_i}{\partial S / \partial t}=\frac{\int d^3k  |f(\vec{k})|^2 k_i}{\int d^3k  |f(\vec{k})|^2 \omega}
\end{equation}
Noting that the integrand in the denominator is always less than the absolute value
of the integrand in the numerator, i.e.
\begin{equation}
|f(\vec{k})|^2 \omega > |f(\vec{k})|^2 |k_i|
\end{equation}
we conclude that
\begin{equation}
|\frac{dx_i}{dt}|<1
\end{equation}
\par
(b)--Consider the case of 1+1 dimensional solution. The energy and momentum
densities is defined for an ensamble of particles as
\begin{equation}
{\cal E}=-R^2\frac{\partial S}{\partial t}= \frac{1}{2}\int dk dk' f(k) f^*(k') exp[-i(\omega -\omega')t+i(k-k')x] \{\omega +\omega '\}
\end{equation}
\begin{equation}
{\cal P}=R^2\frac{\partial S}{\partial x}= \frac{1}{2}\int dk dk' f(k) f^*(k') exp[-i(\omega -\omega')t+i(k-k')x] \{k+k'\}
\end{equation}
The averages of ${\cal E}$ and ${\cal P}$ over a long time interval $(-T/2,T/2)$
are
\begin{equation}
<{\cal E}>\rightarrow \frac{2 \pi}{T}\int dk |f(k)|^2 \frac{\omega ^2}{k}
\end{equation}
\begin{equation}
<{\cal P}>\rightarrow \frac{2 \pi}{T}\int dk |f(k)|^2 \omega
\end{equation}
If we define the average velocity as the ratio of $<{\cal P}>$ and $<{\cal E}>$ then,
we have
\begin{equation}
\frac{<{\cal P}>}{<{\cal E}>}=\frac{\int dk |f(k)|^2 \omega}{\int dk |f(k)|^2 \frac{\omega ^2}{k}} <1
\end{equation}
\par
The conclusion is that at least under the above conditions (i.e., localization of $f(\vec{k})$
and separations larger than the inverse of bandwidth of $f(\vec{k})$, or for long time intervals),
the observed velocity of the particle is less than that of light.
\newline
\newpage
{\large REFERENCES}
\newline
[1]- Ghose, P. and Home, D., {\it Phys. Lett. A\/}, {\bf 191}, No. 5,6, 362, (1994).
\newline
[2]- Bohm, D. and Hiley, B.J., {\it The Undivided Universe\/}, page 286, Routledge (1993).
\newline
[3]- Holland,P. R., {\it The Quantum Theory of Motion\/}, Cambridge University Press, London/New York, (1993).
\newline
\end{bf}
\end{document}